%Paper: hep-th/9510134
%From: Juan Maldacena <malda@puhep1.Princeton.EDU>
%Date: Wed, 18 Oct 1995 12:26:15 -0400

%%%%%%%%%%%%%%%%%%%%%%%%%%%%%%%%%%%%%%%%%%%%%%%%%%%%%%%%%%%%%%%%%%%%%%%%%%%%%%%
\input harvmac.tex
\input epsf.tex

\noblackbox

\def\sq2{\sqrt{2}}
\def\ra{\rightarrow}

\def\p{\partial}

\def\bp{\bar \p}
%--------+---------+---------+---------+---------+---------+---------+
\lref\ber{ K. Behrndt, HUB-EP-95 [hep-th 9506106]. }
\lref\buscher{ T. Buscher, Phys. Lett. B201 (1988) 466. }
\lref\dhoker{ E. D'Hoker and D. Phong, Rev. Mod. Phys. 60 (1988)
  917. }
\lref\dhstring{A. Dabholkar and J. A. Harvey, Phys. Rev. Lett. 63
  (1989) 478; A. Dabholkar, G.W. Gibbons, J. A. Harvey and F. Ruiz
  Ruiz, Nucl. Phys. B340 (1990) 33.}
\lref\duff{M.J. Duff and J. Rahmfeld, Phys. Lett. B345 (1995) 441
  [hep-th/9406105]; see also M.J. Duff, R.R. Khuri and J.X. Liu,
  Physics Reports 259 (1995) 213 [hep-th/9412184] and references
  therein.}
\lref\ferrel{R. Ferrell and D. Eardley, Phys. Rev. Lett. 59 (1987)
  1617.}
\lref\garfinkle{ D. Garfinkle, Phys. Rev. D46 (1992) 4286.}
\lref\gibbonstown{ G.W. Gibbons and P.K. Townsend, Phys. Lett. B356
  (1995) 472 [hep-th/9506131]. }
\lref\harvey{ J.P. Gauntlett, J.A. Harvey, M.M. Robinson, and
  D. Waldram, Nucl.Phys. B411 (1994) 461 [hep-th/9305066]. }
\lref\hor{G.T. Horowitz and A.A. Tseytlin, Phys. Rev. D51 (1995)
  2896 [hep-th/9409021].}
\lref\horsen{ G. Horowitz and A. Sen, UCSBTH-95-27, TIFR/TH/95-46
  [hep-th/9509108].}
\lref\hortsesing{ G.T. Horowitz and A.A. Tseytlin, Phys. Rev. D50
  (1994) 5204 [hep-th/9406067].}
\lref\hortseprl{G.T. Horowitz and A.A. Tseytlin, Phys. Rev. Lett.
  73 (1994) 3351 [hep-th/9408040].}
\lref\kallosh{R. Kallosh and A. Linde, SU-ITP-95-14 [hep-th/9507022];
  R. Kallosh, D. Kastor, T. Ort{\'\i}n and T. Torma,
  Phys. Rev. D50 (1994) 6374 [hep-th/9406059];
  E. Bergshoeff, R. Kallosh and T. Ort{\'\i}n, Phys. Rev. D50
  (1994) 5188 [hep-th/9406009].}
\lref\kawai{H. Kawai, D.C. Lewellen and S.H.H. Tye,
  Nucl. Phys. B269 (1986) 1.}
\lref\khurimyers{ R. Khuri and C. Myers, McGill/95-38, CERN-TH/95-213
 [hep-th/9508045].}
\lref\khuriscat{R. Khuri, Nucl. Phys. B376 (1992) 350;
  A. Felce and T. Samols, Phys. Lett. B308 (1993) 30 [hep-th/9211118].}
\lref\landaumech{L. Landau and E. Lifshitz, {\it  Mechanics},
  Third Edition, Pergamon Press.}
\lref\landautctof{L. Landau and E. Lifshitz, {\it The Classical Theory
  of Fields}, Fourth Edition, Pergamon Press.}
\lref\mantongib{N. Manton, Phys. Lett. 154B (1985) 397;
  G. Gibbons and N. Manton, DAMTP 95-29 [hep-th/9506052].}
\lref\peet{A. Peet, PUPT-1548 (to appear in Nucl. Phys. B)
  [hep-th/9506200].}
\lref\senbh{A. Sen, Nucl. Phys. B440 (1995) 421 [hep-th/9411187].}
\lref\senentropy{A. Sen, TIFR-TH-95-19 [hep-th/9504147].}
\lref\senstring{A. Sen, Nucl. Phys. B388 (1992) 457 [hep-th/9206016].}
\lref\sentendfourd{ A. Sen, Int. J. Mod. Phys. A9 (1994) 3707
  [hep-th/9402002].}
\lref\shiraishi{K. Shiraishi, Nucl. Phys. B402 (1993) 399.}
\lref\sm{J. Maharana and J. Schwarz, Nucl. Phys. B 396 (1993) 3
  [hep-th/9207016].}
\lref\stu{L. Susskind, L. Thorlacius and J. Uglum, Phys. Rev. D48
  (1993) 3743 [hep-th/9306069].}
\lref\tsesch{ A.S. Schwartz and A.A. Tseytlin, Nucl. Phys. B399 (1993)
  691 [hep-th/9210015].}
\lref\tseytlin{A. Tseytlin, Phys. Lett. B176 (1986) 92.}
\lref\tseytlinaas{A.A. Tseytlin, IMPERIAL/TP/94-95/54
  [hep-th/9509050].}
\lref\wilczek{  C. Holzhey and F. Wilczek, Nucl. Phys. B380 (1992)
  447 [hep-th/9202014].}
\lref\wstring{D. Waldram, Phys. Rev. D47 (1993) 2528
  [hep-th/9210031].}
\lref\sensixd{A. Sen, Nucl. Phys. B450 (1995) 103 [hep-th/9504027].}
\lref\harveystrom{ J. Harvey and A. Strominger, Nucl. Phys. B449
(1995) 535 [hep-th/9504047]. }
\lref\mirjam{M. Cveti\v{c} and A. Tseytlin, IASSNS-HEP-95/79,
Imperial/TP/95-96/4 [hep-th/9510097] }

%--------+---------+---------+---------+---------+---------+---------+
\Title{\vbox{\baselineskip12pt
\hbox{PUPT-1565}\hbox{hep-th/9510134}}}
{\vbox{\centerline{\bf Extremal Black Holes As Fundamental Strings}}}
\centerline{Curtis G. Callan, Jr.,
  Juan M. Maldacena and Amanda W. Peet\footnote{$^\dagger$}
{e-mail addresses: callan, malda, peet@puhep1.princeton.edu }
}
\centerline{\it Department of Physics, Princeton University}
\centerline{\it Princeton, NJ 08544}
\vskip.3in
\centerline{\bf Abstract}

We show that polarization dependent string-string scattering provides
new evidence for the identification of the Dabholkar-Harvey (DH)
string solution with the heterotic string itself. First, we construct
excited versions of the DH solution which carry arbitrary left-moving
waves yet are annihilated by half the supersymmetries. These solutions
correspond in a natural way to Bogomolny-bound-saturating excitations
of the ground state of the heterotic string. When the excited string
solutions are compactified to four dimensions, they reduce to Sen's
family of extremal black hole solutions of the toroidally compactified
heterotic string. We then study the large impact parameter scattering
of two such string solutions.  We develop methods which go beyond the
metric on moduli space approximation and allow us to read off the
subleading polarization dependent scattering amplitudes. We find
perfect agreement with heterotic string tree amplitude predictions for
the scattering of corresponding string states.  Taken together, these
results clearly identify the string states responsible for Sen's
extremal black hole entropy. We end with a brief discussion of
implications for the black hole information problem.

\smallskip

\Date{10/95}
\eject
%--------+---------+---------+---------+---------+---------+---------+
\newsec{Introduction}

It is an old and attractive idea that the very massive string states
must be black holes because their Schwarzschild radii are bigger than
their Compton wavelengths. This identification is most compelling for
the special case of extremal supersymmetric black holes: they have
vanishing Bekenstein-Hawking entropy and therefore can reasonably be
thought of, as string states usually are, as ordinary particles.  Sen
\senbh\ has found a family of such solutions to the six dimensionally
compactified low energy heterotic string effective action.  Their
charge/mass relation is identical to that of a fundamental string in
its right moving ground state but in an arbitrary left moving state.
Like such string states, they preserve two of the four supersymmetries
and saturate Bogomolny bounds. As a result, the lowest order mass
formulas are protected from corrections which might disturb the
correspondence of mass spectra.  In \duff\ this correspondence between
Sen's extremal black holes and string states was made explicit at the
level of the quantum numbers.

There is, however, a small paradox.  On the one hand, the mass of the
extremal string states is determined by the charges which in turn are
related to the {\it total} ~left moving oscillator level $N_L$, and
the degeneracy at fixed $N_L$ gives rise to an entropy which increases
with increasing $N_L$.  On the other hand, the horizon area of the
black hole solutions is zero and so is the Bekenstein-Hawking
entropy. Sen was able to resolve this discrepancy by showing
\senentropy\ that the area of the ``stretched horizon'' \stu\ gives an
entropy which matches the left moving oscillator degeneracy. But, if
the degenerate substates of an extremal black hole can be identified
with specific string states, it should be possible to use string
theory to study the dynamics of these otherwise hidden states.
That is the goal of this paper.

We study black holes in lower dimensional compactified target spaces
by building them out of extended string like solutions in ten
dimensional target space.  The well known Dabholkar-Harvey (DH)
solutions \dhstring\ are of this type, but not sufficiently general
for our purposes, so we must first extend our knowledge of these
solutions in several directions. To begin with, we construct a new
class of solutions, generalizing the DH solution to the case of
arbitrary left moving excitation. Just as the DH solution should
correspond to the $N_R=1/2, N_L=1$ state of a winding heterotic
string, so the new solutions correspond to its $N_R=1/2, N_L>1$
excitations. They are constructed along the lines of past work on
``plane-fronted waves'' \refs{\hor,\hortsesing} but describe a wider
class of stringy excitations than have been explicitly considered
before.  It has already been remarked that certain lower dimensional
black holes can be constructed by compactifying the ten dimensional DH
fundamental string. We will show that Sen's complete set of four
dimensional extremal black holes can be obtained by compactifying our
excited generalizations of the DH string.

In an investigation of string dynamics we compute the classical large
impact parameter scattering of two parallel oscillating strings.  The
effect of the scattering event on the transverse excitations allows us
to read off a ``polarization dependent'' scattering amplitude.  We
compare these results to the predictions of the heterotic string for
the polarization dependent scattering of the fundamental string states
and find exact agreement. This considerably strengthens previous
evidence from polarization independent scattering on the one hand for
identity of the DH soliton and the heterotic string
\refs{\khuriscat,\harvey} and, on the other hand, for the identity of
certain four dimensional extremal black holes and the compactified
heterotic string \khurimyers.

In addition, we have an identification of the four dimensional black
holes with specific compactified states of the excited ($N_L>0$)
heterotic string itself.  This raises the expectation that questions
of the dynamics within the degenerate state space associated with
black hole entropy can then be addressed directly, by using string
results.  We will comment on this possibility and its relevance to the
more general problems of black hole information loss.

The paper is organized as follows: In section two we construct and
discuss the solutions that generalize the DH string solution.  We also
show how to construct black hole solutions of the compactified theory
by superposing the excited DH string solutions, and show that the
complete set of previously known static black hole solutions can be so
obtained.  In section three we calculate the classical low velocity,
small angle scattering of these string solutions, paying particular
attention to polarization dependence. In section four we calculate the
string theory amplitudes for the corresponding string states in the
same limit, i.e. small angle and velocity, and demonstrate agreement
with the classical result.  The last section contains a discussion of
possible consequences of the black hole/string identification and our
conclusions.

%--------+---------+---------+---------+---------+---------+---------+
\newsec{Classical Strings in $d=10$ and Black Holes in $d<10$}
%--------+---------+---------+---------+---------+---------+---------+
\subsec{Static strings}

We will begin by reviewing the ten dimensional fundamental string
solution of Dabholkar and Harvey \dhstring, a singular solution of
$N=1$ ten dimensional supergravity that preserves half of the
spacetime supersymmetries. This solution was subsequently generalized
in \refs{\wstring,\senstring} to include charge and momentum flowing
along the string. The action is
$$ S_{10} = \int d^{10}x \sqrt{-G} e^{-\Phi} [ R +
\nabla_\mu \Phi \nabla^\mu \Phi
-{1\over 12} H_{\mu\nu\lambda} H^{\mu\nu\lambda}
- 2 F_{\mu\nu}^I F^{I~\mu\nu} ]
$$
where the gauge fields $F_{\mu\nu}^I$ are in the $U(1)^{16}$ subgroup
of $E_8\times E_8$. For an extended string pointing in the direction
$\hat 9$, the nontrivial spacetime fields of the generalized DH string
are\footnote{$^\dagger$}{We use conventions for the fields and ten
dimensional action as in \sentendfourd. For the constants we take
$\alpha^\prime=2$ and put the asymptotic value of the string coupling
constant to be $g=1$, so we have $G_N^{2/(d-2)}=2$ for the Newton
constant.}
$$
(G_{\alpha\beta}) =
e^{\Phi} \pmatrix{ -(1+C) & C \cr C & 1-C }
$$
\eqn\tend{
B_{09} =- e^\Phi +1  ~~~~~~~~~~~~~~ A^I_0= - A^I_9 = N^I e^\Phi
}
where $\alpha,\beta\in (0,9)$
\eqn\cee{
C = R + 2 e^\Phi N^I N^I
}
and $e^{-\Phi}, R, N^I $ are harmonic functions of the eight
transverse coordinates (e.g.  $\partial_i
\partial_i e^{-\Phi} =0,~~i=1,\ldots,8 $).  For a string of mass,
momentum, and sixteen $U(1)$ charges per unit length equal to
$m,p,q^I_L$ these functions have the form
\eqn\lamb{
e^{-\Phi}= 1 +2 m \Lambda ~~~~~~~~~~~R= - 2 p \Lambda ~~~~~~~~~~~~~
N^I = {q^I_L \over \sqrt{2}} ~\Lambda}
with
\eqn\lambdaeq{
\Lambda = { 8 \over \pi^3 r^6 } \equiv { c_9 \over r^6}
  ~~~~~~~~~~~~ r^2= x^2_1 +
  \cdots x^2_8 .  }
As is usual with supersymmetric solutions, there are multiple string
solutions, where the harmonic functions become superpositions, e.g.
$$
m\Lambda = \sum_a {m_a c_9 \over |{\vec r} - {\vec r}_a|^6}
$$
with the $a$--th string located at transverse position $\vec r_a$.

It has always seemed more than plausible that these solutions should
be identified with states of the fundamental string carrying various
amounts of zero-mode momentum. On the other hand, in string theory
these momenta must satisfy the $L_0-\tilde L_0=0$ level-matching
condition, while no conditions are imposed on the corresponding
parameters $m,p,q^I_L$ of the classical solution. We think we see how
to resolve this mystery: Although the classical solutions are singular
at $r=0$ and are not solutions there in the strict sense unless a
source is provided at the singularity, one can make the singularity
invisible to outside observers by imposing one condition on the
momenta.  To see this, consider a light ray moving towards the string.
Its trajectory is described at small $r$ by
$$
0 = [-w + {\cal O}({1\over \Lambda}) ]dt^2 + 2 w dx^9 dt +
[ -w +  {\cal O}({1\over \Lambda}) ]dx^9 dx^9
+ dx^i dx^i
$$
where $w = w(m,p,q^I_L) = -p/m + q^I_L q^I_L/(4m^2)$ is a constant
and $\Lambda$ blows up as $r^{-6}$ as $r \rightarrow 0$.  Thus
$$
\left| {dx^i\over dt} \right| = \sqrt{w(m,p,q^I_L)
\left(1 - {d x^9 \over d t} \right)^2
+ {\cal O}\left({1\over\Lambda}\right) }
$$
and the time taken for a light ray to escape from $r=0$ to a finite
distance diverges if and only if
\eqn\matching{
w(m,p,q^I_L)=0\qquad {\rm or}\qquad  4mp = q^I_Lq^I_L
}
We will see later that this is identical to the level matching
condition for the fundamental string with corresponding zero-mode
momenta and with oscillator excitations $N_R=1/2, N_L=1$. One might
have expected a correspondence with the strict string ground state,
but it is not too surprising to see a typical normal-ordering
shift. When we construct classical solutions with excitations
corresponding to strings with $N_L\gg 1$ we will again find that the
condition for the singularity to be invisible is the same as the
string level-matching condition. By itself, this seems to us a
significant piece of evidence for the identity of the string solutions
with the fundamental string.

%--------+---------+---------+---------+---------+---------+---------+
\subsec{Oscillating strings}

Lower dimensional black holes with entropy have quantum numbers which
are conjectured to correspond to string states with arbitrary left
moving oscillator excitation.  This motivates us to generalize the
static string solutions of the previous section to solutions carrying
propagating transverse waves.  In fact, we will be able to construct
solutions that represent multiple parallel oscillating strings, each
string carrying a different traveling wave.

We start by constructing the explicit solution for one oscillating
string.  It was not apparent to us how to generalize the static string
solutions so we started instead with the chiral null models considered
in \hor\ and elsewhere.  We will soon see that a particular chiral null
model gives a generalization of the static string which has transverse
excitations.  In the natural light-cone coordinates
$u=x^9-x^0,~v=x^9+x^0$ the particular solution of interest to us takes
the form
\eqn\single{\eqalign{
ds^2 =& e^{ \Phi(r)} du [ dv + K(r) du + 2 f'^i(u) dx^i ]
+ dx^i dx^i
\cr B_{uv} = & e^{\Phi(r)}~~~~~~~~~~~~~~ B_{ui} = 2 e^{\Phi(r)}
f'^i(u)
}
}
where $f^i(u)$ are arbitrary functions describing a traveling wave on
the string, $e^{-\Phi}$ and $K$ are harmonic functions, and in ten
dimensions $ e^{-\Phi(r)} = 1 + { 2 m c_9 / r^6}$ and
$K(r)=2pc_9/r^6$.  Since this metric is not manifestly asymptotically
flat, we prefer to make the simple change of coordinates
\eqn\change{
x^i= y^i - {f^i(u) } ~~~~~~~~~~~v=\tilde v +   \int^u
{ [f'^i(u_0) ]^2 du_0  }
}
which puts the fields in the form
\eqn\singleasy{\eqalign{
ds^2 =& e^{ \Phi(\vec r,u)} du \left[
d\tilde v -2 (e^{-\Phi(\vec r,u)} -1) f'^i(u) dy^i \right. + \cr
  & \left. +
  \left( (e^{-\Phi(\vec r,u)} -1) [f'^i(u)]^2 + K(\vec r,u) \right)
  du \right] + dy^i dy^i \cr
B_{uv} = & e^{ \Phi(\vec r,u)}~~~~~~~~~~~~~~
B_{ui} = 2 e^{ \Phi(\vec r,u) } f'^i(u) }
}
where $\Phi(\vec r,u)=\Phi(\vec r-\vec f(u))$ and $K(\vec r,u)= K(\vec
r-\vec f(u))$. The metric is now manifestly asymptotically flat, and,
in the limit $f^i(u)\to 0$, it reduces to the static solution \tend\
with zero charge.

Now we turn to the problem of constructing multiple oscillating string
solutions. It turns out that the following simple generalization of
the structure of the fields of the single string gives us an ansatz of
sufficient generality:
\eqn\mult{\eqalign{
ds^2 =& F(\vec r,u) du[dv +
K(\vec r,u)du + 2 V_i(\vec r,u) dx^i] + dx^i dx^i \cr
B_{uv} = & F(\vec r,u) ~~~~~~~~~~~~~~
 B_{ui} = 2 F(\vec r,u)  V_i(\vec r,u) ~~~~~~~~~\Phi=\Phi(\vec r,u)~.
}}
All that we have really done is to generalize the chiral null models
of \hor\ by allowing $u$ dependence in $F$.  We first demand that this
configuration, regarded as a background of heterotic string theory,
preserve half of the spacetime supersymmetries. The supersymmetry
variations of the fermionic fields are
\eqn\vari{\eqalign{
\delta_\epsilon \lambda =& {\sq2 \over 4 \kappa }[
-{1\over 2} \gamma^\mu \p_\mu \Phi + {1\over 12} H_{\mu \nu \rho}
\gamma^{\mu \nu \rho} ]\epsilon
\cr
\delta_\epsilon\psi_\mu = &{1\over \kappa }[ \p_\mu + {1\over 4}
(\omega_\mu^{\hat\nu \hat\rho} -
{1\over 2}H_\mu^{~{\hat\nu \hat\rho}} )
 \Gamma_{{\hat\nu \hat\rho}}]\epsilon
\cr
\delta_\epsilon \chi =& -{1\over 4 g}
F_{\mu\nu}\Gamma^{\mu\nu}\epsilon
}}
where $\kappa =\sqrt{8\pi G_N}$ is the gravitational coupling
constant, plain greek letters label coordinate indices, and letters
with a hat label tangent space indices. Coordinate and tangent indices
are related by the zehnbeins $e^{\hat\nu}_\mu$ and
$\omega_\mu^{{\hat\nu \hat\rho}} $ is the corresponding spin
connection.  $\Gamma_{\hat\nu}$ are the flat space gamma matrices
satisfying $ \{ \Gamma_{\hat\nu} , \Gamma_{ \hat\rho} \} = 2
\eta_{{\hat\nu \hat\rho}}$, $ \gamma^{\mu } = e^\mu_{\hat\nu}
\Gamma^{\hat\nu} $ and $\gamma^{\mu_1\cdots \mu_n}$ is the
antisymmetrized product with unit weight (i.e. dividing by the number
of terms).  Choosing the zehnbein
$$ e_\mu^{~\hat\nu} = \pmatrix{ F^{1/2}&0&0\cr
-{F^{3/2} V^2 } & F^{1/2} & {F V_i } \cr
0&0 & 1_8 }
$$
and demanding that the variations \vari\ vanish we find
\eqn\conditions{\eqalign{
&\log F(\vec r,u) = \Phi(\vec r ,u) + z(u) \cr &\epsilon = F^{1/4}
\epsilon_0,~~~~~~~~~~~ ~~~~~~~ \Gamma_{\hat v} \epsilon_0 =0 }
}
where $\epsilon_0 $ is a constant spinor and $z(u) $ is an arbitrary
function.

The next step is to impose the equations of motion
$$
R^{[-]}_{\mu\nu} + D^{[-]}_{\mu } D^{[-]}_{\nu } \Phi =0
$$
The notation indicates that the curvature and covariant derivatives
are constructed out of the generalized connection
$ \Gamma^{[-]\mu}_{\nu\delta} =
\Gamma^\mu_{\nu\delta} - {1\over 2} H^\mu_{\nu\delta}$. These
equations summarize the beta functions for both the metric and the
antisymmetric tensor. They yield the following additional conditions:
\eqn\equations{\eqalign{
&\p_i\p_i e^{-\Phi} =0\cr
 & \p_j\p_j V_i - \p_i \p_j V_j +
e^{- z } \p_i \p_u e^{-\Phi} = 0\cr
& \p_u \p_i V_i - \p_u( e^{-z} \p_u  e^{-\Phi } )
- {1\over 2} \p_i \p_i K = 0
}}
These Laplacian equations can easily be solved and the solution
of interest to us is
\eqn\multstr{\eqalign{
e^{-\Phi } =& ~ 1 + \sum_a 2 m_a \Lambda( \vec r - \vec f_a(u) )\cr
V_i=& - \sum_a 2 m_a  f'^i_a(u) \Lambda( \vec r - \vec f_a(u) ) \cr
K =&  \sum_a 2 p_a(u)\Lambda(\vec r - \vec f_a(u)) \cr
z =& 0 ~~~~~~~~~~~~~~~~~~ \Lambda(\vec r) = c_9/|\vec r|^6
} }
where $f_a^i(u) $ is an arbitrary function. This clearly represents a
collection of oscillating strings with an arbitrary traveling wave,
specified by $f_a^i(u)$, on the $a$--th string. What is surprising and
remarkable is that this non-static solution preserves half the
supersymmetries and is therefore presumably a BPS-saturated state.
Note that this construction of oscillating strings works equally well
for any dimension $d > 4 $, with $\Lambda=c_{d-1}/|\vec r|^{d-4}$.

In string theory, the left moving oscillators associated with the
gauge degrees of freedom can also be excited and we would like to find
a corresponding classical solution. It turns out that we can find a
supersymmetric solution having an oscillating current by adding to
\multstr\ and \mult\ a set of $U(1)$ gauge fields defined as in
\tend\ with
\eqn\gauge{
N^I(\vec r,u)=  \sum_a {q^I_{L,a}(u)\over\sqrt{2}}
\Lambda(\vec r- \vec f_a(u))
}
and by augmenting the function $K$ as follows:
\eqn\kay{
K(\vec r,u) = \sum_a 2 p_a(u)\Lambda(\vec r - \vec f_a(u))
- 2 e^{\Phi(\vec r,u)} N^I(\vec r,u) N^I(\vec r,u)
}
$K$ thus takes the same form as does $[-C]$ in \tend ,
now with $u$ dependent $p$ and $q^I_{L,a}$
$$\eqalign{
p_a(u) =& ~ p_a + \tilde p_a(u) \cr
q^I_{L,a}(u) =& ~ q^I_{L,a} + \tilde q^I_{L,a}(u) }
$$
In these equations the tildes denote the oscillating parts while the
$p_a,q^I_{L,a}$ denote the constant, zero mode pieces.

With the gauge fields given by \gauge, the function $K$ as in \kay,
and the dilaton and $V_i$'s as in \multstr, we see that we have
achieved a multiple string generalization of the previously known
static string.  This new solution has oscillations in the transverse
directions, and oscillating densities for the longitudinal momentum,
charge and current.

There remains the question whether these solutions correspond to exact
conformal field theories. In \hor, Horowitz and Tseytlin demonstrate
that what amounts to the single string version of our solution is
conformal to all orders in $\alpha^\prime$. In the appendix we present
a slight generalization of their argument which, we believe, extends
their proof to our multiple oscillating string solution.

Just as for the single static string, there is the question whether
one can restrict the parameters so as to make the singularities at the
center of each string invisible to outside observers. We will discuss
this issue using the same strategy we applied to the single static
string: the trajectory of a light ray moving perpendicular to an
oscillating string near its singularity is governed by the equation
$$
 \left(dx^i\over dt\right)^2 + 2 { dx^i \over dt }
f'^i \left(1-{dx^9 \over dt } \right)  -
w\left(1-{dx^9 \over dt } \right)^2 -
{\cal O}\left({1\over \Lambda }\right) =0
$$
where $w = -p(u)/m + q^I_L(u) q^I_L(u)/(4m^2)$.  The
$(t,\vec r)$ cross term in the trajectory equation comes from the
functions $V_i$ and is not present for the static string.
Rewriting this as
$$
\left| {d\over dt} [\vec r - \vec f(u)] \right|
= \sqrt{[w+(\vec f')^2]
\left(1-{dx^9 \over dt } \right)^2  +
{\cal O}\left({1\over \Lambda(\vec r - \vec f(u)) }\right)
}
$$
we see that it takes a light ray an infinite time to escape from the
singularity, thereby indicating that the singularity is hidden, if and
only if the condition
\eqn\flevelmatching{
4 p(u) m = [q^I_L(u) q^I_L(u)] + 4 m^2 [f'^i(u) f'^i(u)]
}
is satisfied. We will see in the next section that this ``invisible
singularity'' condition agrees, modulo the normal-ordering subtlety,
with the $L_0=\tilde L_0$ stringy level matching condition with left
moving oscillators excited.

As a final check on our understanding of the physics of these
solutions, we should verify that the excited strings do indeed
transport physical momentum and angular momentum. Since we have
written the metric in a gauge where it approaches the Minkowski metric
at spatial infinity, we can use standard ADM or Bondi mass techniques
to read off kinetic quantities from surface integrals over the
deviations of the metric from Minkowski form.  Following \dhstring, we
pass to the physical (Einstein) metric $g_E = e^{-\Phi/4} g_{string}$,
expand it at infinity as $g_{E \mu\nu} = \eta_{\mu\nu} + h_{\mu\nu}$
and use standard methods to construct conserved quantities from
surface integrals linear in $h_{\mu\nu}$.  Of course the Bondi mass
and its kinetic analogs are most appropriate for this problem since
things in general depend on a light-cone coordinate.  In particular,
for a single oscillating string, we find that the transverse momentum
on a slice of constant $u$ is
\eqn\mom{ P_i = m f'^i(u) }
in precise accord with ``violin string'' intuition about the
kinematics of disturbances on strings. Similarly, the net
longitudinal/time momentum $\Theta_{\alpha\beta}$ in a constant $u$
slice is
$$
(\Theta_{\alpha\beta}) =
\pmatrix{m+P_{u} & -P_{u} \cr -P_{u} & -m+P_{u} \cr}
$$
where
$$
P_{u} = p(u) - 2m f'^i(u) f'^i(u)~.
$$
is the physical longitudinal momentum per unit length. In this
expression we have used the fact that $P_v = {m/2}$.  Finally, we
consider angular momenta. For the string in ten dimensions there are
four independent (spatial) planes and thus four independent angular
momenta $M^{ij}$.  Evaluating as an example $M^{12}$ we obtain
$$
M^{12} = m ( f'^1 f^2 - f'^2 f^1)(u)~.
$$
There are no surprises here, just a useful consistency check.

%--------+---------+---------+---------+---------+---------+---------+
\subsec{$d\leq 9$ Black Holes From $d=10$ Fundamental Strings}

The static multi-string solutions discussed at the beginning of this
section can be used to produce pointlike solutions of the toroidally
compactified theory in lower dimensions: all that is needed is to
place the centers of the strings on a lattice in the transverse space
and to compactify the $\hat 9$ direction on a circle. To be more
precise, we build a periodic $(9-d)$ dimensional array of strings in
the $\hat d$ to $\hat 8$ directions by taking $\Lambda_d =
\sum_{Lattice}
\Lambda({\vec r}-{\vec r}_a)$. For large $(d-1)$ dimensional spatial
distances $\rho$ we can ignore the dependence on the internal
dimensions and find, to leading order in large $\rho$,
\eqn\lambdaddimbh{
\Lambda_d = {c_d \over \rho^{d-3}} \qquad {\rm where}\quad
  c_d = {{16\pi}\over{[(d-3)\omega_{d-2}]}}
}
and $\omega_d$ is the area of the $d$ dimensional unit sphere. We
could have taken directly $\Lambda = c_d/\rho^{d-3}$ as a solution of
the Laplace equation in the uncompactified dimensions, but this
obscures the essential connection to underlying string states. As we
will now show, the result of this proceedure can be interpreted as a
lower-dimensional extremal black hole. The general idea that
ten-dimensional string solutions can be used to generate
four-dimensional black holes is not new and has been explored in
\refs{\hortseprl,\ber,\kallosh,\mirjam}.
Our contribution will be a more precise understanding of the relation
between black hole parameters and string quantum numbers.

We now look in more detail at the $d$-dimensional fields generated by
this compactification. Using the dimensional reduction procedure of
\sm\ and the conventions of \sentendfourd, we find that the
$d$-dimensional dilaton is
\eqn\ddimdil{
e^{-2 \Phi_d } = e^{-2\Phi_{10}} (G_{99})
= { e^{-\Phi_{10} } ( 1-C )}
= {1 + { 2(m+p) \Lambda_d } + {( 4 m p - q_L^2 )\Lambda_d^2}}~.
}
This and all the other fields obtained by dimensional reduction turn
out to be identical to those of Sen's four-dimensional black holes and
their higher-dimensional generalizations \refs{\senbh,\peet} with the
parameter identifications
\eqn\mqrql{\eqalign{
M_{ADM} = & ~~(m+p)\cr
[{\omega_{d-2}\over 4\pi}] {1\over 2\sq2}
Q_R^a =& ~~(m+p) \delta^{a 1}
\cr
[{\omega_{d-2}\over 4\pi}] {1\over 2\sq2}
Q_L^a =& ~~(m-p) \delta^{a 1} + q_L \delta^{a I}
}
}
where $a$ is any index from $1,\ldots (10-d)$ and $I$ is any index
from $(11-d)\ldots (26-d)$, and both $a$ and $I$ are $U(1)$ group
labels. These are not completely generic charges since the first six
components of $Q_L$ are pointing in the same direction as $Q_R$.  A
black hole with generic charges can be obtained by boosting the
previous solution \tend\ along an internal direction and then reducing
back to $d$ dimensions.

We noted earlier that in order to hide the singularity of the static
string solution, the densities $m$, $p$ and $q_L$ had to satisfy the
level-matching constraint \matching. In terms of the parameters of the
lower-dimensional black holes \mqrql, this condition reads
$$
\vec Q_R^2 = \vec Q_L^2~.
$$
{}From the field theory point of view there is, however, nothing wrong
with the lower-dimensional black holes of \refs{\senbh,\peet} which do
not satisfy this condition: they are still perfectly supersymmetric
and extremal. Moreover, these black holes have entropy and explanation
of this entropy was one of the main goals of our work.  It is a rather
natural guess that the missing black holes arise from toroidally
compactifying strings which are in the right moving ground state and
hence annihilated by half the supersymmetries, but in an excited state
of the left moving oscillators. Such string states satisfy the
level-matching condition
\eqn\prlnl{
p_R^2 - p^2_L = 2 (N_L -1)
}
and ought to yield, upon compactification, black holes with
$\vec Q_R^2 \neq \vec Q_L^2$.

This suggests that we study the lower dimensional black holes arising
from compactification of the oscillating string solutions \multstr.
More precisely, those solutions presumably correspond to string states
where the oscillators are in a coherent state with macroscopic
expectation value for $N_L$. One slight puzzle is that these string
solutions oscillate, while the black hole solutions are static. Of
course, all fields are periodic in $u= x^9-x^0$, because $x^9$ is
compactified, and, for small compactification radius, one might argue
that the oscillations should ``average out'', leaving an effective
static solution. We can make this argument more precise: We can
superimpose any collection of solutions of the type \multstr, in
particular a collection of oscillating solutions which differ in the
phase of oscillation but are otherwise identical.  Mathematically this
corresponds to replacing $u\ra u+u_0$ and integrating over $u_0$. This
is a well defined ``averaging'' procedure that is guaranteed to give a
supersymmetric, static solution.

When we further compactify to $d$ dimensions, if the oscillations
point in the internal dimensions we do indeed obtain Sen's most
general black hole solutions \refs{\peet,\senentropy}.  In addition,
the degrees of freedom giving rise to the entropy of the black holes
are very clear in our construction: they are precisely those of the
ten dimensional string.  When the oscillators point in the external
uncompactified directions we expect to get rotating black holes.  Let
us now review the import of the invisible singularity condition for
the oscillating string which gives rise to the static (nonrotating)
black holes. After integrating the ten dimensional fields over $u$ and
expressing the charges and momenta in terms of the $Q_{L,R}$ of the
lower-dimensional black hole, the condition \flevelmatching\ becomes
$$
{1\over 8} [{\omega_{d-2}\over 4\pi}]^2 (\vec Q_R^2 - \vec Q_L^2)
= 2 m^2 \langle (\vec f')^2 \rangle +
{1\over 2}\langle (\tilde q_L^I)^2\rangle
$$
This, together with \mqrql and \prlnl, leads to the natural
interpretation that the left moving oscillator level for the string
state corresponding to the static black hole in $d$ dimensions is
related to the mean square oscillations in ten dimensions by
\eqn\nlitofqt{
N_L = m^2 \langle (\vec f')^2 \rangle + {1\over 4}\langle (\tilde
q_L^I)^2\rangle
}
(quantum normal ordering effects, to which this classical computation
should not be sensitive, would replace $N_L$ by $N_L-1$). Looking at
\nlitofqt, we see that the left moving oscillator level comes from both
macroscopic oscillations associated to the spacetime directions and
charge oscillations associated to the internal directions.  In the
next section, we will see that the same condition arises from quantum
level matching.

%--------+---------+---------+---------+---------+---------+---------+
\newsec{Classical Scattering of Fundamental Strings}

In this section we will study the classical scattering of fundamental
strings in ten dimensions. From this calculation we will also extract
information about the scattering of objects in lower dimensions that
can be constructed from the fundamental string. This includes all
extremal electrically charged black holes in the compactified theory.
The calculation of the scattering will be done in the low velocity and
small angle approximation. The level of approximation used will be the
``test string'' approximation, where we consider one string moving in
the background of the other. This approximation is well justified if
the scattering angle is small, since it amounts to a neglect of the
recoil of the background string.

%--------+---------+---------+---------+---------+---------+---------+
\subsec{Scattering of generalized fundamental strings in $d=10$.}

We consider a background string as in \tend\ with parameters $m,p$ and
$q$. Again, we take the gauge fields to be in a $U(1)^{16}$ subgroup
of $E_8\times E_8$. It will be convenient to bosonize the 32 left
moving fermions of the heterotic string in terms of 16 bosons $
\varphi^I $, $I=1,\ldots ,16$. To obtain the worldsheet action in
terms of the left moving bosons and the right moving fermions, we
start with the standard worldsheet action written in terms of left and
right fermions.  After bosonization, as is usual, anomaly terms which
would arise at one loop for the chiral fermions arise at the classical
level for the bosons.  One also finds that the antisymmetric tensor
must transform under gauge transformations. The precise form of the
worldsheet action is found by demanding gauge invariance and
supersymmetry. We find that
\eqn\action{\eqalign{
S=&- {1 \over 2 \pi } \int d\tau d\sigma
( \sqrt{-h} h^{mn} + \epsilon^{mn} ) \left\{
\p_m X^\mu \p_n X^\nu ( G_{\mu\nu} + 2 A^I_\mu A^I_\nu
+ B_{\mu\nu} ) +\right.\cr
& \left.
 +\p_m \varphi^I \p_n \varphi^I  - 2 \sqrt{2} \p_m X^\mu A_\mu^I
\p_n \varphi^I
+ {i\over 2}
\psi^\mu  \gamma_m {\cal D}_n  \psi^\nu G_{\mu\nu}
 \right\}
}}
where
\eqn\covar{\eqalign{
{\cal D}_{m  \nu}^{\mu} =& \delta^{\mu}_\nu \p_m  +
(\Gamma^\mu_{\delta \nu} + {1\over 2} H^\mu_{\delta \nu} ) \p_m X^\delta
+\sqrt{2} F^{I\mu}_{~~~\nu} (\p_m \varphi^I -\sqrt{2} A_\delta^I
\p_m X^\delta ) \cr
H_{\mu\nu\rho} =& ( \p_\mu B_{\nu\rho} - 2 A^I_\mu F^I_{\nu\rho} ) +
{ \rm ~~ (cyclic~ permutations)}~.
}
}
It will be important to know the form of the supercurrent constraint:
\eqn\scurrent{
 ( \sqrt{-h} h^{mn} -\epsilon^{mn} ) \gamma^r \gamma_n \psi^\mu
\p_r X^\nu G_{\mu\nu}(X) = 0~.
}

Our aim here is to generalize the results of \harvey\ to include
the response of fermionic and oscillator degrees of freedom carried
by the string to small angle scattering. We first make the gauge choice
$$
h_{\tau \sigma } =  0 ~~~~~~~~~~
h_{\tau \tau} =  h_{\tau \tau}(\tau) ~~~~~~~~~~
h_{\sigma \sigma} = h_{\sigma \sigma}(\tau)~.
$$
Our general strategy, following \harvey, is to reduce the worldsheet
(string) action an effective worldline (particle) action by choosing a
restrictive ansatz for the worldsheet fields which reflects the fact
that most degrees of freedom do not actively participate in a low
velocity scattering event. A sufficiently general field configuration
is
\eqn\gauge{\eqalign{
X^0 = &X^0(\tau) + X_L^0(\sigma^-,\tau) \cr
X^i = &X^i(\tau) + X_L^i(\sigma^-,\tau) \cr
X^9= &m'  \sigma + X^9(\tau)+ X_L^9(\sigma^-,\tau) \cr
\varphi^I = & {q^I \over 2} \sigma^- + \varphi^I_L(\sigma^-,\tau) \cr
\psi^\mu = & \psi^\mu (\tau)
} }
where $ \sigma^- = \int^\tau e d\tau - \sigma $ with $e =
\sqrt{-h_{\tau\tau}/h_{\sigma \sigma} } $. The $X^\mu(\tau)$
describe the motion of the string center of mass. The functions
$ X^\mu_L,\varphi^I_L $ describe the state of the left moving
oscillators. They are periodic functions of $\sigma^-$ and we take
them to be slowly varying functions of the second argument $\tau$
in order to allow for possible adiabatic changes in the state
of left-moving oscillation in the course of a soft collision.
In what follows, we assume that this variation is negligible
and will later verify that this is self-consistent. The right moving
fermion $\psi$ is taken to be a worldsheet zero mode (i.e. a function
of $\tau$ only) in order to focus on the dynamics of BPS-saturated
states. Again, we have to verify that higher fermionic modes are not
excited during a collision.

The left moving oscillators influence the center of mass motion only
through the stress energy constraint, a helpful simplification. The
stress energy tensor would classically be set equal to zero, but, as
shown in \harvey, to properly account for quantum zero point energies
and, more subtly, the difference between left and right moving sectors
in the heterotic string, one must instead set $T_{mn} = \pmatrix{
e^2/2 & -e/2 \cr -e/2 & 1/2 }$. This constraint is more conveniently
recast as
\eqn\tconstr{
 \widetilde  T_{mn} = {(N_L - 1)\over 2}
\pmatrix{  - e^2 &  e \cr  e & -1 } ~,}
where $\widetilde T_{mn} $ is the limit of the energy momentum tensor
when the left moving oscillators $ X_L^\mu $ are set to zero and
$$
N_L =  2 \int {d \sigma \over \pi } \left( \p_- X_L^\mu
\p_- X_L^\nu G_{\mu \nu} + \p_- \varphi^I_L\p_- \varphi^I_L
\right)
$$
is what would normally be identified as the left-moving oscillator
level.  For an oscillating string of the kind studied in the previous
section, the expression for $N_L$ reduces to \nlitofqt , as promised.
Note that \tconstr ~amounts to imposing only the zero mode component
of the Virasoro constraints, which is what is necessary now that we
are ignoring the left moving oscillators in the action.

Under the above assumptions, we can reduce the worldsheet action
\action\ to a worldline action for the center of mass coordinates of the
string. The essential point is that, after integrating over $\sigma$
and applying the energy momentum constraint \tconstr, the contribution
of the left-moving oscillators to the action collapses to $ -(N_L-1) e
$ (where we include the quantum normal-ordering constant). The
resulting effective worldline action is
\eqn\redaction{ \eqalign{
S =& -{1\over 2} \int d \tau \left\{ -  e^{-1} \dot{X}^2  + e X'^2
 -  (N_L-1) e +
2 \dot X^\mu X'^9 B_{\mu 9} + \right. \cr
  & \left.
+ 2 \sqrt{2} \p_- \varphi^I (\dot{X^\mu} + e X'^\mu) A_\mu^I
 - {i \over 2 } e^{-1/2} \bar{\psi}^\mu ({\cal D}_0 - e
 {\cal D}_1 )
 ( \gamma_{\hat 0}
 + \gamma_{\hat 1} ) \psi^\nu G_{\mu \nu } \right\}~.
} }
The oscillator level $N_L$ will be a constant of the motion and
strings with left-moving excitations will be just as easy to deal with
as ground state strings. The term with ${\cal D}_1 $ contributes only
through the term in the connection involving $X'^9$. We have also
found it convenient to make the redefinition $ \psi \ra (-h)^{1/8}
\psi$.

Now let us think about solving this action for the motion of one
string in the background of another. We will first work in the crude
leading approximation adequate to study the motion of the center of
mass coordinates. Later we will worry about the more subtle effects
that give polarization-dependent terms in the scattering. As the
Lagrangian is independent of $X^9$, we can eliminate $X^9$ from the
action in favor of a conserved momentum $p' \equiv P_9$ which will
include some fermionic contributions. The value of $p'$ is determined
by the off-diagonal part of \tconstr. Next, we use the equation of
motion for $e$ to eliminate $e$ from the action \redaction\ and the
supercurrent \scurrent. In addition we choose the gauge $X^0(\tau)
=\tau$.  We then expand the action to second order in the velocity for
the bosonic terms, to first order in velocity for the fermionic terms
and keep only terms quadratic in the fermions. With this, the
supercurrent constraint becomes
$$
\psi_0 +\psi_9 + v^i \psi^i =0 \qquad
            {\rm where}\qquad v^i=\dot X^i(\tau)~.
$$
We can use it to replace $\psi_0 $ in the action, eliminating $\psi_9
$ as well. We finally obtain an action for unconstrained center of
mass coordinates and their associated right moving fermions:
\eqn\final{\eqalign{
 S=& \int  d \tau\left\{ {1\over 2} g v^2 -{i\over 2} \left \{
 g  \psi^i \partial_0 \psi^i +
\psi^i \psi^j ( v^i\p_j g - v^j \p_i g ) \right\}\right\} \cr
g =& M'  + (2 m' p + 2 m p'- q^I q'^I ) \Lambda
}}
where $M' = m' + p' $, $\Lambda $ is as in \lambdaeq, and the primed
and unprimed quantities refer to the test string and the background
string respectively.  The function $g$ is (proportional to) the metric
on moduli space and governs the low-energy scattering in the usual
way. It depends explicitly on the charge and momentum parameters of
the strings and implicitly on the left-moving oscillator level through
the level-matching constraints. By a slight generalization, we can
derive the worldline action for a string moving in the background of
an arbitrary superposition of fundamental strings.  There is an
asymmetry in \final ~between the two strings which is due to the fact
that we fixed the position of the background string. A symmetric
expression can be found by demanding that it be symmetric in the two
strings, invariant under translations, and that it reduces to \final
{}~when we fix the position of one of the two strings. For the bosonic
part of the action, this prescription gives
\eqn\morefinal{
S = \int {1\over 2} M v^2_1  +  {1\over 2} M' v^2_2 +
(2 m'p + 2 m p' - q^I q'^I ) \Lambda(r_1 - r_2) {1\over 2} (v_1 -v_2)^2
}

For the moment we will ignore the fermionic terms in \final , and
calculate the small angle scattering cross section due to the metric
on moduli space. We could infer it from the classical relation between
scattering angle and impact parameter or we could quantize the
Lagrangian \morefinal\ and compute the Born approximation. The two
results are different because quantum effects make the classical limit
unreliable for small scattering angles in dimensions $d \ge 5$: the
uncertainty principle prevents the relative fluctuations in scattering
angle and impact parameter from being simultaneously small. The Born
approximation is, however, reliable in all dimensions when applied to
small angle low momentum transfer scattering. We apply it to
\morefinal\ by taking the $r$-dependent part (the part proportional to
$\Lambda$) as the perturbation and find
$$ {d\sigma\over d\Omega} \sim
 { \Delta_{12}^2 \over \theta^4 }
$$
where $\Delta_{12} = 2 m' p + 2 m p'- q^I q'^I $. The feature that the
cross-section depends on angle but not on energy is a typical metric
on moduli space result.

The result agrees with previous calculations of string-string
scattering, such as \harvey, and generalizes them by evaluating the
metric on moduli space for the most general string in its right moving
ground state. In this regard, it may be helpful to use the
level-matching conditions to rewrite $\Delta_{12}$ in terms of the
left moving oscillator levels:
$$
\Delta_{12} = {1\over 2}[(N_L-1)+(N'_L-1)+(q_L-q'_L)^2]
$$
This makes it clear that for identical $N_L>1$ strings, the Born
approximation cross section goes as
\eqn\crossbb{
{d\sigma\over d\Omega}_{\{N_L>1\}} \sim {(N_L-1)^2\over\theta^4}~.
}
When $N_L=1$, as is true for the fundamental string solution of
\dhstring\ or the ``$a=\sqrt{3}$'' extremal charged dilaton black
hole, the metric on moduli space is flat and the worldline action must
be expanded to one more order in $v^2$ to extract the cross section.
When do such an expansion, we find the improved action
\eqn\nlone{
 S_{\{N_L=1\}} = {1\over 2} \int d \tau \{ M v^2 + { M \over 4 }
v^4( 1 + 2 M   \Lambda + q^I q^I \Lambda^2 ) \}
}
(where $M = m+p$ and $ 4mp = q^I q^I $ because of the $N_L=1$
condition) and a Born cross section
\eqn\crossbbnone{
{d\sigma\over d\Omega}_{\{N_L=1\}} \sim {M^2 v^4\over\theta^4}
}
Doubts had been expressed in the literature on whether radiation terms
might not be of comparable order to the $O(v^4)$ potential terms
responsible for the scattering. We estimate the effective forces due
to radiation in an identical particle system to be of still higher
order than $v^4$, hence negligible. Indeed, it is shown in
\landautctof\ that in the case of identical particles interacting
through gravity one can find a reliable two-body Lagrangian to order
$v^4$.

%--------+---------+---------+---------+---------+---------+---------+
\subsec{Polarization dependence of the scattering.}

We now want to go a step beyond the metric on moduli space
approximation and consider the effect of scattering on the
``polarization'' state of the strings. The heterotic string has both
right moving and left moving polarization degrees of freedom. The
right moving ones are described by the fermions $\psi^i$ that have
already appeared in the action \final\ and it is quite easy to discuss
their scattering dynamics. The left movers are more subtle and we will
discuss their behavior separately.

The right moving degrees of freedom always remain microscopic, in the
sense that, even for large mass strings, their excitation number
remains finite and there are no long range fields associated with
them.  They are analogous to the spin degrees of freedom of a massive
particle.  To describe their dynamics, we just have to quantize the
full Lagrangian \final, including the $\psi^i$. After the rescaling
$\psi \ra g^{-1/2}\psi$, the $\psi^i$ have the standard
anticommutation relations of $\gamma$ matrices. The Hilbert space is
spanned by eight-dimensional spinors and describes the Ramond sector
of the heterotic string, as we can also see from \gauge . To get the
right physics, we have to project onto a definite $S0(8)$ chirality of
the spinors. In the Neveu Schwarz sector, the fermions are
antiperiodic in the $\sigma$ direction, and we change the ansatz in
\gauge\ for picking out the fermion ground state to
$\psi^\mu(\tau,\sigma) = ( e^{i \sigma}
\psi^\mu(\tau) + {\rm c.c.})/\sqrt{2}$.

We will now concentrate on the NS sector. The reduction to a worldline
action gives something similar to \final\ but with complex fermions
$\psi^i(\tau)$.  The NS fermions behave as fermionic creation and
annihilation operators and the lowest allowed states have a
single creation operator acting on the vacuum (as usual, the vacuum
itself is eliminated by the GSO projection). The action becomes
$$\eqalign{
S=& \int  d \tau  \left( L_{free} + L_{int} \right)\cr
L_{int} =& {1\over 2} v^2 \Delta_{12} \Lambda - i
 { \Delta_{12} \over M' }  (v^i\p_j \Lambda -v^j\p_i\Lambda )
 \psi^{i\dagger} \psi^j }
$$
where the fermions are normalized so that $\{\psi^{i\dagger},\psi^j\}=
\delta_{ij}$ and we have kept in the fermionic term only the leading
power of $1/r^6$ since we are interested in small angles. Representing
the states by $\eta^i \psi^\dagger_i |0\rangle $ we can calculate the
Born amplitude
\eqn\poldep{
{\cal A} =  \eta^i_1 \eta^i_3  \Delta_{12} {k^2 \over 2}
 {\cal F}_q(c_9 /r^6) + \eta^i_3
( q_i k_j - k_j q_i )  \eta^j_1 {\cal F}_q(c_9 /r^6)
}
where ${\cal F}_q(c_9/r^6)$ is the Fourier transform of $c_9/r^6$ with
respect to the transfer momentum $q$. Here $\eta_1, \eta_3$ denote the
polarizations of the initial and final states.  If the initial and
final polarizations are equal we get the same Born amplitude we found
in the previous bosonic calculation:
$$
{\cal A}_{no~flip} = { \Delta_{12} \over \theta^2 }
$$

In order to make a concrete calculation of polarization changing
amplitudes, let us consider two strings with equal masses and charges:
one being a background string sitting at the origin and the other
string coming toward it with small velocity $\vec v = v \hat 2$ and
with a large impact parameter $b$ in the direction $\hat 3$, as
depicted in Fig. 1.
\vskip 1cm
\vbox{
{\centerline{~~~~~
{\epsfxsize=4.5in \epsfbox{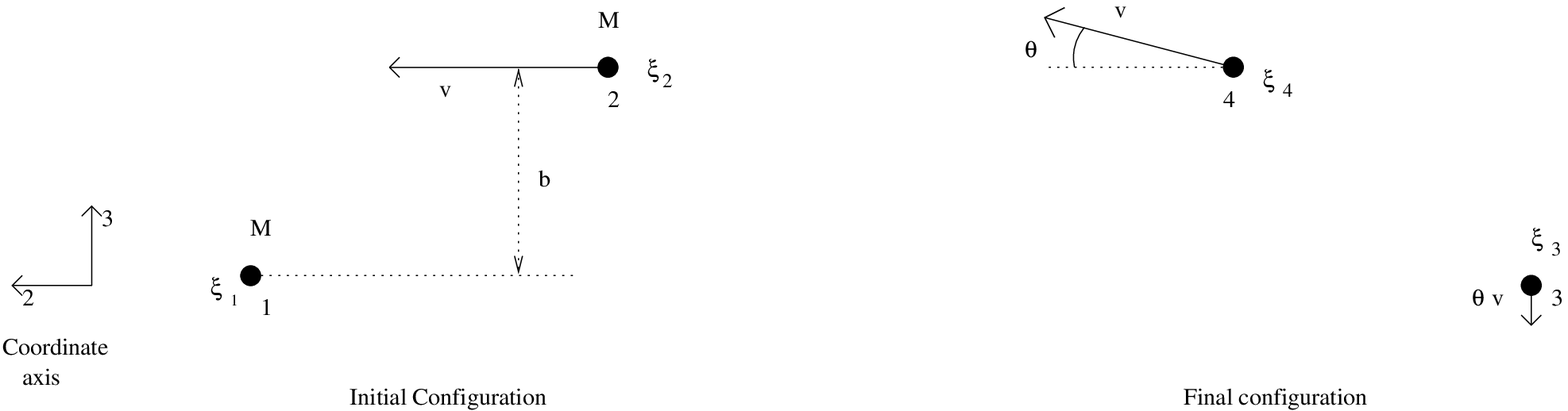}  }}}
{\centerline{
Figure 1: Scattering configuration. One string is initially
at rest and the other}}
{\centerline{  has  velocity $v$ and impact parameter
$b$ and is scattered by a small  angle $\theta$.
}}
}
\vskip .5cm
\noindent
For this situation we have for small angles $\vec q=M'v\theta~\hat 3$
.  The amplitude for a polarization to rotate from the direction $\hat
3$ to $\hat 2$ can then be read off from \poldep:
$$
{\cal A}_{32}= 2 { \Delta_{12}  \over \theta }
$$
The crucial point is that, apart from kinematic factors, the same
potential governs both the spin-flip and no-flip Born amplitude. A
more concise statement of the result is
\eqn\rightflip{
{{{\cal A}_{flip}}\over{{\cal A}_{no flip}}} = 2\theta
}
The probability that a polarization originally in the direction $\hat
3$ would flip in the course of scattering through angle $\theta$ and
wind up pointing in the direction $\hat 2$ is $P_{flip} = ( 2 \theta
)^2$.  The calculation is essentially the same for the Ramond case.

At this point we would like to comment on the expectation that, for
solitons breaking some of the target space supersymmetries, the full
action in moduli space should possess a worldline supersymmetry.  The
action \final\ is not supersymmetric. This is not a contradiction
because we are treating the Neveu-Schwarz and Ramond sectors on a
different footing. The supersymmetries would transform one sector into
the other.

We now turn to the question of the left moving polarization
dependence.  Unlike the right moving polarizations, which are
analogous to semiclassical spin degrees of freedom, these
polarizations can have a macroscopic effect since the energy in these
modes can be comparable with the total mass. Indeed, we saw in
previous sections how to construct classical solutions carrying these
polarization degrees of freedom. In principle we could obtain a low
energy action which includes the left moving oscillators by doing a
more accurate derivation of \final, much as we did for the fermions.
Instead of doing this we will start with the classical equations of
motion of the full worldsheet action \action\ and project out linear
decoupled effective equations of motion for the left moving
oscillators.  This is only possible in the small-angle scattering
limit, of course.  It would be nice to have a more unified treatment
of the different kinds of polarization degrees of freedom, but our ad
hoc approach is adequate for our current purposes.

For the analysis, it is more convenient to choose a coordinate system
in which the test string is sitting at the origin and the background
string is moving as in Fig. 1 (rather than the other way around).  Now
we allow the background string to have any nonzero $m,p,q^I$.  The
equations of motion for the test string in conformal gauge are
\eqn\motion{\eqalign{
\p_+\p_-X^\mu =& -(\Gamma^\mu_{~\nu \delta} -{1\over 2}
 H^\mu_{~\nu \delta} ) \p_+X^\nu
\p_-X^\delta  + \cr
+ & \sqrt{2} F^{\mu I}_{~\nu} \p_+X^\nu
(\p_-\varphi^I - \sqrt{2} A^I_\delta \p_- X^\delta )\cr
0 = & \p_- (\p_+\varphi^I -\sqrt{2}  A_{\mu}^I \p_+X^\mu )
}}
where $H$ is as defined in \covar .
We write the coordinates of the string as
\eqn\coord{
\eqalign{
X^0 =& (m'+p')\tau +X^0_L(\sigma^-,\tau) \cr
X^9 =& m'\sigma +p'\tau +X^9_L(\sigma^-,\tau) \cr
X^i =& X^i_L(\sigma^-,\tau) \cr
\varphi^I =& {q^I\over 2} \sigma^-  +\varphi^I_L(\sigma^-,\tau)
}}
where $\sigma^\pm = \tau \pm \sigma $ are the light cone coordinates
on the string world sheet.  In the above equations, $X_L$ and
$\varphi_L$ are periodic functions of $\sigma^-$ and slowly varying
functions of the second argument $\tau$.  This explicit $\tau$
dependence is very small ($\p_\tau X^\mu_L <<\p_{\sigma^- } X^\mu_L $)
and is caused by the time dependence of the background.  We also
assume that the oscillation amplitude is small compared to the
characteristic scale over which the background fields vary, which can
be achieved if the amplitude is much smaller than the distance between
the two strings.  These approximations enable us to replace in the
background fields the average values of the coordinates, so that the
equation \motion\ becomes linear in $ X^\mu_L $.  When we replace
\coord ~in \motion ~we get a $\sigma$ independent part which
represents the velocity dependent total force on the string.  This
force is responsible for the no-polarization-flip scattering described
above. The sigma-dependent part gives a set of equations for the
evolution of the left-moving oscillators:
\eqn\motionleft{\eqalign{
 \p_\tau (\p_- X_L^\mu ) = (m'+p')&\left\{
-[\Gamma^\mu_{~0\delta} -{1\over 2}
 H^\mu_{~0\delta} + \Gamma^\mu_{~9\delta}
 -{1\over 2} H^\mu_{~9\delta} ]
\p_- X_L^\delta  + \right. \cr
& \left.
+\sqrt{2} (F^{\mu I}_{~0} + F^{\mu I}_{~9} ) (\p_-
 \varphi^I_L -\sqrt{2} A_\delta^I \p_- X^\delta )  \right\}
}}
and an analogous equation for $\varphi^I_L $.  The right hand side of
\motionleft\ vanishes identically for a static background string.
When we boost it to get the expression for the moving string the
result is no longer zero and can be expanded in powers of the
velocity.  The overall effect of the background string on the
oscillators is a change in the direction of oscillation. This change
is independent of the frequency of the oscillator because we can
Fourier transform in $\sigma_-$ both sides of \motionleft.
Replacing in \motionleft\ the values of the backgound fields for a
moving string  in \motionleft, we find
\eqn\motion{
\p_{X^0} \p_- X^\mu_L = {v^2 \over 2 } (  p_L^\mu \p_\nu \Lambda
 -  p_{L\nu}\p^\mu \Lambda  ) \p_- X^\nu_L
}
where we kept only the leading terms in $1/r^6$ and $p^\mu_L$ is the
left moving transverse momentum of the background string.

Since the equations of motion for the left moving oscillators are
linear, we can read off the interaction Hamiltonian from \motion .
We introduce creation and annihilation operators for the transverse
oscillators that satisfy $ [a^{\mu \dagger}_n,a^\nu_m ] =
\eta^{\mu\nu}\delta_{nm} $.  The interaction Hamiltonian then becomes
\eqn\hint{
H_{int} = -i {v^2 \over 2} (  p_{L \mu}\p_\nu \Lambda
  -  p_{L \nu} \p_\mu \Lambda) \sum_{n>0} a^{\mu \dagger}_n a^\nu_n
}
This leads to a Born amplitude
$$ \langle \chi_{final} |  { v^2 } {\cal F}_q(c_9/r^6)
( q_\mu  p_L^\nu - q^\nu p_{L \mu} ) \sum_{n>0} a^{\nu \dagger}_n
a^\mu_n | \chi_{initial} \rangle
$$
where $\chi_{initial,final}$ are the initial and final oscillator
states. They satisfy the usual Virasoro physical state conditions. The
evolution dictated by \hint\ preserves them.  In fact, it can be seen
that the terms proportional to the longitudinal momentum in \hint\
change the polarizations in such a way that the physical state
conditions are satisfied.  We see that in this approximation at most
one oscillator can change its state.  In order to take a specific
example consider the case in which an oscillator that is originally
pointing in the direction $\hat 3$ flips to the direction $\hat
2$. The ratio of polarization changing to polarization preserving
amplitudes is
\eqn\pflip{
{ {\cal A}_{flip}
\over {\cal A}_{no~flip} }
 =  {v^2 M^2 \theta \over  \Delta_{12} }
}

We can now justify an important simplifying assumption introduced at
the beginning of this section: We assumed in writing \redaction\ that
terms leading to changes in the left-moving oscillator state could be
neglected, at least in the leading ``metric on moduli space"
approximation to the scattering. The calculation we have just
completed, showing the vanishing of the polarization flip probability
as $v \ra 0$, obviously justifies that initial assumption.

We note finally that if we are scattering two {\it identical } $N_L=1$
strings this ratio becomes undefined because of the vanishing
denominator and a more accurate calculation must be done. We omit the
details, but the result is $2 \theta$ just as it was for the right
moving fermions, as one would expect. All these more or less
complicated polarization dependences are not so much interesting for
themselves as for the way they will compare with direct string theory
calculations of the same quantities.

%--------+---------+---------+---------+---------+---------+---------+
\subsec{Lower dimensional objects}

The next thing we would like to do is to investigate scattering
involving lower dimensional black holes.  To begin with, let us
consider two body black hole scattering.  We have seen that $d$
dimensional black holes in the compactified theories can be thought of
as string arrays in the uncompactified ten dimensional theory.  If we
scatter two such objects, then we can imagine doing a ``test array''
approximation, where we consider one array moving in the field of the
other. This idea actually works, for two reasons.  First of all, each
string in the test array moves independently of its array-mates, since
there are no static forces between the strings of a given array by
virtue of supersymmetry. The Lagrangian for the motion of each of
these strings reduces, in the low velocity limit, to
\final\ but with the function $\Lambda$ being that of the
other array of strings. The second important point is that there are
only two body forces in this Lagrangian, since the interaction term is
a sum of the interaction terms of the test string with each of the
background strings, and since the Lagrangian \final\ represents only
two body forces. Note that, in general, one might have expected to
find higher-body forces; in fact, for four dimensional extremal
Reissner- N\"{o}rdstrom black holes there are up to four body
interactions \ferrel.

The Lagrangian \final\ thus enables us to calculate once and for all
the two body scattering cross section for charged extremal black holes
in $d$ dimensions, in the low velocity and small angle approximation.
Before writing the final Lagrangian, we make a check of the static
forces between $d$ dimensional extremal black holes.  We find that, in
order for the static force to vanish, the right moving charges of both
black holes have to be parallel: $ Q_R= \lambda Q'_R $ with $ \lambda
>0 $.  Then we have that the low velocity Lagrangian is
$$
 S = \int {1\over 2}  v^2 \left[M' +
\left({\omega_{d-2} \over 8 \sqrt{2} \pi }\right)^2 {
( \vec Q_R \cdot \vec Q'_R - \vec Q_L \cdot \vec Q'_L ) c_{d-3}
 \over   \rho^{d-3} }
\right]
$$
for a $d$ dimensional black hole.

This Lagrangian can in fact also be obtained directly in $d$
dimensions by using the test particle approximation, and it seems to
us that this agreement is another consistency check of the
correspondence we have found between our classical ten dimensional
strings and lower dimensional static black holes.

We see from this Lagrangian that the metric on the lower dimensional
black hole moduli space is of the same form as for the string in ten
dimensions, except for the power of the distance. The cross section is
then the same as in ten dimensions, up to an overall constant which is
independent of any of the parameters of the black holes.

The metric on moduli space is zero only in the case where we have two
identical $N_L=1$ black holes ($N_L=0$ black holes have $ Q_L > Q_R =
2 \sqrt{2} M $ and contain naked singularities).  In this case,
performing a $O(10-d)\times O(26-d)$ rotation and a subsequent
redefinition of the dilaton, we can reduce the system to a pair of
$a=\sqrt{3}$ dilaton black holes, whose moduli space metric was shown
to be flat \refs{\shiraishi,\khurimyers,\gibbonstown}. Using the
Lagrangian \nlone\ for identical $N_L=1$ strings, we find the first
nonvanishing term in the black hole lagrangian in this case by simply
replacing $\Lambda$ in \nlone\ with its value for a $d$ dimensional
black hole \lambdaddimbh.  The cross section is then again the same as
for ten dimensions, up to the same overall constant as for the $N_L>1$
case.

Another calculation that we wish to do is the scattering of massless
particles off a black hole.  This is again not so interesting in
itself as for later comparison with the string theory answer.  To do
this calculation classically, we consider small perturbations of these
massless fields around the black hole background and expand the action
to second order.  Obviously, the first order perturbations vanish
because the black hole background is a solution of the equations of
motion, so we have that
$$
S[\phi^i_0 + \delta \phi^i] \sim S[\phi^i_0 ] + {1\over 2}
{\delta^2 S \over \delta \phi^i  \delta \phi^j } \delta \phi^i
 \delta\phi^j
$$
where $\phi^i$ denotes all the fields appearing in the full $d$
dimensional action. Using the $O(10-d)\times O(26-d)$ invariance of
the $d$ dimensional action and the asymptotic conditions \refs{\senbh}
on the fields, we can assume that the right and left moving charges
are parallel and pointing in the internal direction $\hat 1$: $Q^a_R =
|Q_R| \delta^{a 1}$ and $Q^a_L = |Q_L| \delta^{a 1} $.

At this point we can divide the fields into the ones that are excited
in the black hole background (the dilaton, the metric, some moduli and
some gauge fields) and the ones that are not (the antisymmetric
tensor, and the other moduli and gauge fields). Let the quadratic
action including both types of fields be denoted
$$
S_2 = S_{exc} + S_{rest}
$$
We'll concentrate on $S_{rest}$ since it is much the simpler of the
two.  Its explicit form is
\eqn\rest{
\eqalign{
 S_{rest} =& {1 \over 32 \pi } \int d^4 x \sqrt{-g}[ -{1\over 12}
e^{-2 \Phi } H_{\mu\nu\rho} H^{\mu\nu\rho} - \cr
&- e^{-\Phi}
\sum_{a,b \not = 1,d+1} F_{\mu\nu}^{(a)} (L{\cal M}L)_{ab} F^{(b)\mu\nu} +
 {1\over 8} Tr'(\p_\mu{\cal M} L \p^\mu {\cal M} L) ]
}}
where the metric and dilaton are those of the black hole background.
$Tr'$ is the usual trace, excluding the elements of the moduli matrix
${\cal M}_{ab}$ with $a,b \in (1,d+1)$.  The reason for this index
structure is not important for our discussion but it can be found
easily from the form of the black hole solutions in
\refs{\senbh,\peet}.

We can now concentrate on the propagation of the antisymmetric tensor
$B_{\mu\nu}$ or initially unexcited gauge fields $A_\mu$ or scalars
${\cal M}$, but for simplicity let us analyze the propagation of some
element $\zeta$ of the scalar field matrix ${\cal M}$. The action will
be
\eqn\scalar{
S = { 1 \over 256 \pi} \int d^d x \{ -e^{ 2 \Phi_d} \p_0 \zeta
 \p_0 \zeta +
\p_i \zeta \p_i \zeta \}
}
and so the field equation is
$$
e^{ 2 \Phi_d} \p_0^2 \zeta - \p_i^2 \zeta = 0
$$
Assuming a time dependence $\zeta(t,\vec x) = e^{-i\omega t}
\zeta(\vec x)$, and plugging in the black hole dilaton field
dependence of \ddimdil, we get a Schr\"{o}dinger equation with
potential
\eqn\pot{
V = \omega^2 (1-e^{-2 \Phi_d } )=
 - { 2 M c_d \omega^2 \over |x|^{d-3} } -
 \left({\omega_{d-2} \over 8 \sqrt{2} \pi }\right)^2
{ (Q_R^2 - Q_L^2 )
 \omega^2 c_d^2 \over
|x|^{2(d-3)}  }~,
}
leading to the following small angle scattering cross section:
\eqn\massless{
 {d \sigma \over d\Omega } \sim {  M^2 \over \theta^4}~.
}

The simple form of the potential \pot\ can also be used to illuminate
a previously known curiousity involving $d$ dimensional black
holes. It is a familiar fact about the Schr\"odinger equation that
sufficiently singular attractive potentials can lead to violations of
unitarity in the form of partial or complete absorption of incoming
waves. In the black hole context, however, ``absorption'' just
corresponds to falling through the event horizon and is perfectly
acceptable. The potential \pot\ has different consequences in this
respect in dimensions $d=4$ and $d>4$. For $d>4$, the potential is so
singular that there is a finite absorption probability for any
energy. For $d=4$, the potential is marginally singular and there is
absorption only if the potential is stronger than a critical value. To
be precise, a particle is completely reflected if its energy $\omega$
is less than a ``mass-gap'' defined by
$$
(Q_R^2 - Q_L^2 ) \omega^2 < 1/8~.
$$
This phenomenon had already been noted in the particular example of
the four dimensional $a=1$ dilaton black hole, in which case $Q_L=0$
and $Q_R$ is essentially the black hole mass \wilczek.
\footnote{$^\dagger$}{A curious fact is
that when one inspects the supersymmetric part of the heterotic string
tree amplitude for our kinematic situation, one sees that for $Q_L=0$
the above threshold corresponds precisely to the energy necessary to
excite the next mass level of the free string. This fact seems to be
peculiar to four dimensions, however.}  Note that in the case
$|Q_L|\ge |Q_R|$ the quadratic term in \pot\ is absent or repulsive
and no modes, regardless of their energy, are absorbed. This is again
consistent with the result in \wilczek\ that the mass gap is infinite
for four dimensional dilatonic black holes with $ a > 1 $ (
$a=\sqrt{3} $ in this case)\footnote{$^\ddagger$}{We thank Finn Larsen
for discussions on this issue}.  This mass gap in four dimensions was
used to reconcile the fact that the Hawking temperature of the
extremal black hole is nonzero (!) with the picture of these black
holes as elementary particles.  In $d>4$ the Hawking temperature
vanishes \peet\ and this is consistent with the lack of a mass gap, as
is implied by \pot.

The analysis for the gauge fields and antisymmetric tensor field in
\rest , once the gauge is fixed, gives a similar action to
\scalar\ and the same small angle cross section \massless\ when
we consider equal initial and final polarizations.  The small angle
scattering analysis of the fields in $S_{exc} $ also gives the same
small angle cross section \massless .  This universal cross section,
proportional to the mass of the black hole, corresponds to
gravitational scattering.

%--------+---------+---------+---------+---------+---------+---------+
\newsec{String Tree Amplitude Calculations}

We now turn to the calculation of extended string scattering
amplitudes by string tree amplitude methods. Our goal is to establish
a correspondence between particular states of the heterotic string and
classical string solutions.  Evidence for such a correspondence has
previously been found for ground state strings, but we will be able to
extend it to a much larger class of states and in addition address the
issue of polarization dependence.  The lowest order scattering
amplitudes were compared for a restricted class of four dimensional
black holes and string states in \khurimyers.

The context is the heterotic string with the $X^9$ coordinate
compactified on a macroscopic circle of radius $R_9$. We study strings
which wind once around $X^9$, are in the right-moving ground state
($N_R=1/2$), are in a general state of the left-moving oscillators
($N_L\ge 0$) and carry charge $q^I$ (and current).  Such states are
annihilated by half the supersymmetries and saturate the Bogomolny
bound, since from the right moving point of view they correspond to
massless ten dimensional states.  Depending on whether we are in the
NS or R sector we have bosons or fermions. Bosons of this type are
created by vertex operators like
\eqn\vertex{
V= \xi .\psi_R~p_R.\psi_R (z) {\cal O}_L({\bar z} )
e^{ i [\vec p_R \vec X(z) + \vec p_L \vec X(\bar z) ] }
}
where $\cal O$ is a left-moving operator of weight $N_L$
containing all the left-moving polarizations. The momenta are
$$
p^\mu_{L} = (\hat p^{\hat\mu}, {n \over R_9}- {R_9\over 2}, q^I )
\qquad
p^\mu_{R} = (\hat p^{\hat\mu}, {n \over R_9}+ {R_9\over 2} )
$$
where $\hat p$ is the nine-dimensional momentum and the next entry
gives the momenta in the $X^9$ direction. The on-shell conditions on
the ten-dimensional momenta, $0 = p_R^2 = p_L^2 + 2(N_L-1)$, do two
things: they fix the nine-dimensional mass-squared through the
relation $\hat p^2 = (R_9/2+n/R_9)^2$ and they constrain the total
number of left moving oscillators (both coordinate and gauge) through
$2(N_L-1) = 2 n - q^Iq^I $. A given $N_L$ can usually be achieved by
exciting many different combinations of oscillators and higher mass
levels are increasingly degenerate.

We will study the two-body small-angle scattering of such particles
using the method of \kawai\ to express the tree-level heterotic string
amplitude as the product of an open superstring amplitude for the
right-movers and an open bosonic string amplitude for the left-movers:
\eqn\closed{
{\cal A}^{het} \sim  \sin(\pi u/2) {\cal A}^{open}_{ss}(s,u)
{\cal A}^{open}_{bos}(u,t) ~.
}
These amplitudes are conveniently written in terms of the
ten-dimensional right-moving Mandelstam variables:
$$
s = - (p_R^1 + p_R^2)^2 \qquad t =  - (p_R^1 + p_R^4)^2 \qquad
u=  - (p_R^1 + p_R^3)^2 ~.
$$
We will use the convention that $\sum_i p^{R,L}_i=0$. The
superstring amplitude is simple:
\eqn\super{
{\cal A}^{open}_{ss}(s,u) = {\Gamma(-{s/ 2 }  ) \Gamma( -{u/ 2 } )
\over \Gamma({t/ 2 }  +1  ) }
K_{ss}
}
where $K_{ss}$ is a kinematic factor depending on the right-moving
polarizations and momenta. The terms of direct relevance to us are
\eqn\polss{
 K_{ss} = - {st\over 4}(\xi_1 . \xi_3)(\xi_2 . \xi_4) +
\left[{1\over 2} s (\xi_1 . p_4)(\xi_3 . p_2) +
 {1\over 2} t (\xi_3 . p_4)(\xi_1 . p_2)\right]
(\xi_2 . \xi_4 ) + \dots
}
(all momenta and polarizations here should carry a $R$ label).
The bosonic factor is more complicated because of the left-moving
oscillator vertices ${\cal O}_L$, but its general form can
be worked out following \kawai:
\eqn\open{\eqalign{
 {\cal A}^{open}_{bos}(u,t) =& \sum_{\{n\} } C_{\{n\} }
\int^\infty_1 dx x^{p_R^1.p_R^2 } (x-1)^{p_R^1.p_R^3 }
x^{-\tilde n_{12}} (x-1)^{ -\tilde n_{13} } \cr
 = & \sum_{\{n\} } C_{\{n\} }
 {\Gamma(-{t/ 2 } -1 + \tilde n_{12} + \tilde n_{13}   )
\Gamma( -{u/ 2} - \tilde n_{13} +1 )
\over \Gamma({s/ 2} + \tilde n_{12} ) }
} }
where the ${\tilde n}_{ij} = n_{ij} + p^R_i p^R_j - p^L_i p^L_j$ are
certain integers: the $p^R_i p^R_j - p^L_i p^L_j$ part arises because,
while this Veneziano integral is really built out of $p_L$, we chose
to write it in terms of $p_R$; the $n_{ij} $ come from powers of
$z_i-z_j$ arising from Wick-contracting the ${\cal O}_L(z_i)$ with
each other and with the exponentials; the $ C_{\{n\}}$ contain the
left-moving polarizations hidden in the ${\cal O}_L$.  Mutual locality
of vertex operators in string theory guarantees that the $\tilde n$
are indeed integers.

We will study low velocity, small angle (large impact parameter)
scattering of two of these string states. To keep the formulas simple,
we take both nine-dimensional masses equal, but we let the left-moving
momenta be different.  In the center of mass, the ten-dimensional
momenta can be written
$$\eqalign{
p_R^1 = (M\gamma,M\gamma \vec v , M) &\qquad
 p_R^2 = (M\gamma,-M\gamma \vec v , M) \cr
 p_R^3 = -(M\gamma,M\gamma \vec w , M) &\qquad
 p_R^4 = -(M\gamma, -M\gamma\vec w , M)}
$$
where $\gamma$ is the usual relativistic factor, $ v^2= w^2$ and $\vec
w$ forms a small angle ${\theta_c}$.  We denote by $\theta_c$ the
scattering angle in the center of mass frame and $\theta=\theta_c/2$
is the scattering angle in the rest frame of string one, as defined in
fig 1.  with $\vec v$. The mass $M$ is related to $N_L$ and $R_9$ as
previously described. For small velocity and nearly forward scattering
the Mandelstam invariants are:
$$
s\sim -t \sim 4 M^2 v^2 ~~~~~~~~~~~~~u
\sim  - M^2 v^2 {\theta_c}^2 \ll 1
$$
Note that we are not yet assuming that $s,t \ll 1$.
We are interested in the singularities of \open\
 in the limit $u\to 0$ (large impact parameter).

The structure of \open\ is such that such a $u$-channel pole can only
appear if $\tilde n _{13}=2,1$. We need to find what $C_{n}$ is for
these cases.
 The values of $\tilde n _{13}$ are set by the powers of $(z_1-z_3)$
appearing the {OPE} of the left-moving vertex operators for particle
$1$ (one of the two incoming particles) and particle $3$ (the particle
into which it scatters). The first two terms in this {OPE} are,
somewhat schematically,
$$ {\cal O}_1(z_1) e^{ i p_L^1 X(z_1) }
   {\cal O}_3(z_3) e^{ i p_L^3 X(z_3) } \sim
   {\delta_{13} ~ e^{ i ( p_L^1 + p_L^3 ) X(z_3) } \over
     (z_1-z_3)^{2-u/2}} +
   {{\cal P}(z_3)~e^{ i ( p_L^1 + p_L^3 ) X(z_3)} \over
     (z_1-z_3)^{1-u/2}}
$$
The first term generates contributions to the string tree amplitude
with $\tilde n _{13} = 2$, while the second generates $\tilde n _{13}
= 1$.  There are some constraints which restrict the values of other
$\tilde n_{ij}$'s.  Since the conformal weight of each operator is one
we have $ \sum \tilde{n}_{ij} = 4 $. When we send the point $z_4 \ra
\infty$ to fix the M\"{o}bius invariance of the amplitude and get
\open, we pick the term with
$\tilde{n}_{14}+\tilde{n}_{24}+\tilde{n}_{34}= 2$. This implies that
$\tilde{n}_{12}+\tilde{n}_{23}=2-\tilde n_{13}$.  Using the fact that
$ (p_1+p_3)^{L,R}$ is of order $\theta_c$ we conclude that
$n_{12}+n_{23} = 2-\tilde n_{13}$.  Since $n_{ij} \ge 0 $ we conclude
that $n_{12}=n_{23} =0$ if $\tilde n_{13} =2$, and one of them is $1$
and the other zero if $n_{13} = 1$.

The leading term comes from the identity operator term in the {OPE} of
${\cal O}_1$ with ${\cal O}_3$. If we pick the polarization states
${\cal O}_i$ from an appropriately orthonormalized set, the answer is
either $1$ (polarizations of $1$ and $3$ the same) or $0$
(polarizations of $1$ and $3$ orthogonal), up to terms of order
${\theta_c}$ due to the fact that $p^L_3 $ is not precisely $-p^L_1$.
This term is multiplied by the {OPE} of ${\cal O}_2$ with ${\cal
O}_4$, which yields a Kronecker delta function of the polarizations of
particles $2$ and $4$. The polarization operator ${\cal P}$ appearing
in the next term in the {OPE} will have conformal weight one (one more
than the identity operator in the leading term) and is thus of the
form ${\cal P}\sim \p X_L $ where $X_L$ is one of the 26 leftmoving
bosons.  This $\p X_L$ should be contracted with the exponentials of
the other two vertex operators, since conformal invariance implies
that this three point function has poles of order one when $z_3 \ra
z_2,z_4$.
This contraction will therefore be proportional to some component of
$p^L_2+ p^L_4$ which is of order $\theta$ and which reduces the term's
order in $1/\theta$.  In short, the most divergent term in the
${\theta_c} \to 0$ limit of ${\cal A}^{open}_{bos}\sim 1/{\theta_c}^2$
arises when the polarizations of the scatterers are unchanged. The
normalization of this term is known and is independent of the
polarizations. The same thing is true of the $K_{ss}$ factor in ${\cal
A}^{open}_{ss}$.

We are now ready to put together the net result for the
leading ${\theta_c}\to 0$ behavior of the scattering amplitude with
no polarization flip.
We saw above that $n_{12} = 0 $ so that
$\tilde n_{12}=p^R_1 p^R_2 - p^L_1 p^L_2 \equiv
\Delta_{12}$, which reduces to $N_L-1$ if we are scattering two
identical strings.

With these facts in hand, we can see that
$$\eqalign{
{\cal A}^{open}_{ss} \sim  & {s\over u} \cr
{\cal A}^{open}_{bos} \sim & {1\over u}{ \Gamma( -{t/ 2} +
\Delta_{12} +1 ) \over \Gamma( {s/ 2} +
\Delta_{12} )  }
\sim {1\over u} ( \Delta_{12} +s/2)  \cr
{\cal A}^{het} \sim & u ~{\cal A}^{open}_{ss} {\cal A}^{open}_{bos}
\sim { \Delta_{12} \over {\theta_c}^2 }}
$$
The expression for ${\cal A}^{het}$ is a typical ``metric on
moduli space'' result; the angular
distribution has a well-defined limit as $v\to 0$.
In this limit, for $\Delta_{12} \not = 0$, this implies an
interaction Hamiltonian proportional to $v^2 \Delta_{12} $ in
agreement with \final.
 Somewhat
remarkably, it reproduces the classical string solution  scattering
result \crossbb\ for {\it any} oscillator level (in \harvey, this
result was shown for the lowest level only).
For two identical strings and  $N_L-1=0$, the
leading term vanishes ($\Delta_{12}=0$ and
the metric on moduli space is flat) and we
find
$$
{\cal A}^{het} \sim { M^2 v^2 \over {\theta_c}^2 }
$$
in agreement, once again, with the classical result \crossbbnone.
We can also analyze the case of massless particle scattering
by a black hole. The kinematic configuration is
$$\eqalign{
p_R^1 = (M,\vec 0 , M) &\qquad
 p_R^3 = - (M\gamma, M\gamma \vec v , M) \cr
p_2 = ( E, E \hat x , 0 ) &\qquad p_4 = -(E', E' \hat n, 0 )}
$$
and in the limit $E << M$  we have $E\sim E'$ and
$$
 s\sim -t \sim 2 M E \qquad u = - E^2 {\theta_c}^2
$$
for small ${\theta_c}$. An analysis similar to the above tells us
that the leading small-angle amplitude is diagonal in polarization
and independent of the specific polarization value. The amplitude is
$$ {\cal A}^{het} \sim {M^2 \over {\theta_c}^2} $$
in agreement with \massless .

Now we turn to the polarization flipping amplitudes.
We saw above that all polarization flipping amplitudes are
at least one factor ${\theta_c}$ down with respect to the
polarization preserving ones.
We will analyse the leading case in which the amplitude
goes as $ {\cal A } \sim 1/{\theta_c} $. In this limit
either the right or the left moving polarization changes but
not both.

Let us start with right moving polarization changes.
We see from \polss\ that the polarizations have to be in the
scattering plane. Let us assume, as we did in section 4.2, that
the polarizations are $\xi^R_1 = \hat 3, \xi^R_3  = \hat 2$ and
$\xi_2^R= \xi_4^R$. All right moving polarization dependence of
the amplitude is in the factor \polss\ so that
$$
{ {\cal A}^{flip} \over {\cal A}^{no~flip} } =
{ K_{ss}^{flip} \over K_{ss}^{no flip} } = 2 {\theta}
$$
independent on the value of $N_L$ of both strings, in agreement with
the semiclassical result \rightflip.We can also calculate
the polarization flipping amplitudes for the fermionic
states, which are characterized by a spinor $u_a^R$. We
can verify that also for this case there is an
agreement with the semiclassical calculation.

Now we turn to the left moving oscillators, starting with the
$N_L=1$ case.
For this case the full amplitude involves a polarization
dependent factor $K_{bos}$ which, for the terms we are
interested in, reduces to \polss\ but in terms of the
left moving momenta and Mandelstam variables.
Taking again $ \xi^L_1 = \hat 3, \xi^L_3  = \hat 2$ with
$\xi_2^L= \xi_4^L$ and $p^L_1 \not = p^L_2 $, we have that
$\Delta_{12}$ is nonzero.
We find
$$
{ {\cal A}^{flip} \over {\cal A}^{no~flip} } =
{ K_{bos}^{flip} \over K_{bos}^{no flip} } =
{v^2 M^2 \theta \over  \Delta_{12} }
$$
in agreement with \pflip. If the two strings are identical
we find that this ratio is $2 \theta$, as we found also
in the classical calculation.

We have also found agreement for the case $N_L=2$. The
calculation for higher $N_L$ seems more cumbersome.

%--------+---------+---------+---------+---------+---------+---------+
\newsec{Discussion and Conclusions}

In this work we have studied in some detail multiple classical
oscillating fundamental string solutions of ten dimensional heterotic
string theory which possess mass, $U(1)$ charges, and longitudinal
momentum per unit length.  The oscillations are left moving and the
solutions possess unbroken target space supersymmetry.  Upon toroidal
compactification along with an appropriate averaging procedure, these
solutions yield the full complement of static extremal electrically
charged black holes in lower dimensions previously found in the
literature.  The new solutions which we obtained may include, in
addition, configurations that could have interesting physical
interpretations other than the particular ones explored here.  One
issue that we leave for future work is whether there is a
correspondence between $d$ dimensional BPS-saturated rotating black
holes and string states similar to what we have found for the static
case.  A step in this direction has been taken in \horsen.

In our investigations of the classical solutions we found that the
requirement that curvature singularities be invisible to outside
observers imposed one constraint on the parameters of the classical
solutions.  This invisibility requirement is necessary for any
classical solution to be physically reasonable. Remarkably, this
condition was found to be identical, up to a classically undetectable
normal ordering constant, to the level matching condition for the
fundamental string corresponding to the classical solution.  The
oscillations are characterized by eight transverse arbitrary
functions, corresponding to the eight transverse physical polarization
degrees of freedom of the string states.

It was argued that these multiple oscillating fundamental string
solutions were exact to all orders in the string tension in some
scheme and, since the string coupling does not blow up anywhere for
these electric solutions, loop corrections are expected to be small.
The higher dimensional origin of the $d$ dimensional black holes then
implies that the same things can be said of the classical black hole
solutions.

The relationship between these classical oscillating strings and
fundamental strings was explored via comparison of two-body scattering
amplitudes, in the semiclassical approximation for the oscillating
string solutions and at tree level in string theory.  It was found
that the low velocity small angle scattering cross sections agreed,
and in particular that the lowest order left and right handed
polarization flipping amplitudes agreed in the two different
approaches.  This we regard as additional evidence that the classical
solutions should be regarded as the fundamental strings themselves.
It should be noted that this evidence was dynamical and not simply a
consequence of kinematics.

By using the above direct connection between compactified strings and
black holes, we computed two body black hole scattering in the lower
dimensional theory and found that the result again agrees with that
obtained from string theory.  In an investigation of scattering of
massless particles off a black hole, agreement was also found in the
two different approaches. The leading order dynamics of the black
holes was found to be independent of their internal states.

I has been shown \refs{\senentropy,\peet}
 that the entropy of the $d$ dimensional
extremal black holes, calculated at the ``stretched horizon''
according to the usual area rule, scales in the same way as the entropy
of right-moving ground state strings with increasing total
left moving oscillator level.
In these investigations it was unclear precisely which black
hole degrees of freedom the entropy was counting.  Here, by contrast,
we saw that extremal electrically charged black holes in $d=4\ldots 9$
should be thought of as compactified strings.  Therefore in order to
discover the real internal structure of the $d$ dimensional black
hole, which is a solution of the low energy effective action of string
theory, one should do measurements of the fields with resolution
better than the compactification radius.  At this point, the lower
dimensional effective action description breaks down, and the ten
dimensional nature of the theory becomes manifest.  The correspondence
between the ten dimensional string and the lower dimensional black
hole then makes those degrees of freedom which account for the entropy
manifest.  In this regard, the important length scale is the
compactification scale and not, for example,
$\sqrt{\alpha^\prime}$. In other words, in order to differentiate
between two black holes with the same charges we should measure the
fields with a resolution better than the compactification scale.  On
the other hand, if we insist on a $d$ dimensional description, these
oscillations along the internal directions of
 the ten dimensional fields are viewed as massive
fields in  the $d$ dimensional theory. For a given value of
the charges, different black holes correspond to different ways of
exciting these massive fields.

All of the string states and classical objects studied were
supersymmetric, and under these conditions it is expected that there
are nonrenormalization theorems available to protect the relation
between the mass and certain charges.  In order for the physical
properties of four dimensional black holes to be studied, however, one
needs to know what happens outside the protected enclosure formed by
supersymmetry.  An example would be the excitation spectrum of the
black hole, which we expect to be quite different from the free string
spectrum, because of large gravitational corrections.

Despite the fact that quantum corrections to the classical backgrounds
are formally under control, the physics of the classical string
singularity remains opaque to us.  Possibly the investigation
\tseytlinaas\ may shed light on this issue.  In general, the external
fields of a configuration do not necessarily tell us much about its
real internal structure, but here we have found some evidence that a
classical oscillating string should be thought of as the fundamental
string itself, and thus that BPS-saturated black holes should be
thought of as compactified fundamental strings.  In this context it
would be interesting to explore the issue of pair production of black
holes.  We leave for the future the question of how much more
information may be obtained regarding black holes, for example
non-extremal ones, by using string theory.  It is our hope that the
black hole information problem will be resolved in the context of
string theory.

In the context of string-string duality there is an interesting
application of the solutions we have found.  We have in mind the
duality between the heterotic string on $T^4$ and type IIA on K3. It
is well known that starting from the fundamental string solution of
the form \tend\ in six dimensions ($\Lambda \sim 1/r^2$) and applying
the duality transformation we obtain a non singular soliton solution
of type IIA on K3 \refs{\sensixd,\harveystrom} .  Applying this same
transformation to our oscillating string solution we can construct
oscillating nonsingular soliton string solutions of type IIA on K3.
The condition that the solution be nonsingular in the type IIA theory
is in fact the same as the condition that the singularity be non naked
in the heterotic theory and is the the level matching condition for
heterotic string states.  This establishes the correspondence of type IIA
solitons with the heterotic string beyond the small oscillation zero
mode analysis of \harveystrom.
We intend to explore this possibility further.

%--------+---------+---------+---------+---------+---------+---------+
\newsec{Acknowledgements}

We benefited from discussions with Rajesh Gopakumar, Finn Larsen and
especially Lenny Susskind.

The work of CGC and JMM was supported in part by DOE grant
DE-FG02-91ER40671.  The work of AWP was supported in part by NSF grant
PHY90-21984.

%--------+---------+---------+---------+---------+---------+---------+
\newsec{Appendix: Conformal Invariance}

Here we investigate the model
\eqn\action{\eqalign{
S =& { 1\over \pi  \alpha'} \int d^2z \left\{
F(x,u)\p u[ \bp v + K(x,u)\p u + 2 V_i(x,u) \bp x^i ] \right. \cr
& \left. + \p x_i \p x^i  + { \alpha'\over 8} {\cal R}^{(2)} \Phi(x,u)
\right\}
}}
By doing a coordinate transformation we can set
$K=0$, from now on
we set it to zero.
We will follow closely the method of ref. \hor, generalizing it
to the case when $F$ depends also on $u$. We want to show that
$$ Z(V,U,X) = \int {\cal D}( u,v,x^i) e^{- S - { 1\over \pi
\alpha'} \int d^2z
V(z) \p\bp u + U(z) \p \bp v + X^i(z) \p\bp x^i} $$
is conformally invariant. We have included some source terms for
the fields.
We want to find the conditions for which
this partition function is independent of the conformal
factor of the two dimensional metric $g=e^{ 2 \varphi } \eta$.

We start doing the integration over $v$, which produces
a factor
\eqn\delt{
\delta \left( \bp[ F\p u - \p U ] \right) =
 { \delta( u- u^{cl}) \over  Det[F(x,u^{cl})] Det[ \bp \p - \bp \p_u
F^{-1}(x,u^{cl}) \p U ]
}}
where $u^{cl}$ is the solution to the equation
\eqn\ucl{
 F(x,u^{cl})\p u^{cl} = \p U
}
This equation can be solved iteratively, assuming $U$ small, as
follows
$$ u = u^0 + \sum_{n=1}^\infty ( u^n - u^{n-1} ) $$
$$ u^n = { 1 \over \p } F(x, u^{n-1} )^{-1} \p U $$
where $u^0$ is a constant and $u^n$ contains powers of $U$ up to $U^n$.
$$ u^{cl} = u^0 + { 1 \over \p }  F(x, u^0)^{-1} \p U +
{ 1 \over \p } \p_u F(x, u^0)^{-1} \p U
{ 1 \over \p }  F(x, u^0)^{-1} \p U + \cdots
$$
If we think of these functions in terms of Feynman diagrams, by
replacing the inverse powers of $\p$ by lines, we notice that the
vertices are of the form $ \p^n_u F^{-1}\p U $.  The determinant of
the function $F$ has been calculated in
\buscher \tsesch\ \hortsesing\
\eqn\detf{
{1\over Det[ F] }  = const \times \exp \left[
{ -{1\over 4 \pi} \int d^2z R^{(2)}
(- {1\over 2} \log F )
+{1\over 2\pi} \int d^2 z \bp \log F \p \log F } \right]
}
which has the effect of redefining the dilaton in \action to
$\Phi'=\Phi - \log F $. The second term in the exponential
\detf\ ~can be eliminated by adding to the original action \action\
an order $\alpha'$ counterterm which effectively amounts to
changing the metric to $G_{\mu\nu} \ra G_{\mu\nu}  +
{\alpha' \over 2 } \p_\mu \log F \p_\nu \log F $. We can
view this as a field redefinition (note that $F^2 = \det G $) or
a choice of scheme \tseytlin .
The second determinant in \delt\  can be written introducing a pair
of auxiliary bosonic fields $y^a,~a=1,2$ as
\eqn\detother{\eqalign{
 & Det[ \bp \p - \bp \p_u F^{-1}(x,u^{cl}) \p U ]^{-1}= \cr =& (const.)
\int {\cal D}(y) e^{ - {1 \over \pi \alpha' } \int \{\bp y^a \p y^a -
 \bp y^a \p_u F^{-1}(x,u^{cl}) \p U y^a \}  }
}}
The integration over $u$ can be readily done.
We obtain the effective action
\eqn\seff{\eqalign{
S_{eff} = &
 {1\over \pi \alpha'} \int d^2 z \left\{
\p U 2 V_i(x,u) \bp x^i
- \bp y^a \p_u F^{-1}(x,u^{cl}) \p U y^a \right. + \cr
&
\left. + \p x_i\bp x^i + \p y^a\bp y^a +
 {\alpha' \over 8 }  {\cal R}^{(2)}
\Phi'(x,u) - \bp V F^{-1}(x,u) \p U  + X^i \p\bp x^i
 \right\}
}
}
where we understand that, from now on, where it says $u$ we mean
$u^{cl}$.  Note that we have used \ucl\ to express some of the terms
in \seff.  We have to integrate \seff\ over $x^i$ and $y^a$. The
effective action
\seff\ has, already at the classical level, a dependence
on the conformal factor of the metric, coming from the dilaton term
through $ R^{(2)} = -2 \bp \p \varphi $.  Note that the action \seff\
is non local because of the dependence of $u$ on the other fields of
the theory. We can however expand the action in terms of Feyman
diagrams and analyze the divergences.  This action \seff\ has the
interesting property that all vertices contain a derivative $\p$
acting on a background classical field (e.g. $\p U$ ) so that in a
quantum loop involving $n$ vertices we get an integral which, in the
worst case goes as $ \int d^2p {\bar p^{2 n} / p^{2n +2} } $ which is
not divergent because of its tensor structure (Pauli-Villars
regularization, for example, respects this tensor structure).  We
conclude that the only divergent terms are the ones coming from
tadpole diagrams. When we regularize these divergences we find an
extra dependence on the conformal factor of the metric.  We will use
the heat kernel definitions for contractions at coinciding points
\dhoker\
\eqn\contr{
\langle x(z) x(z) \rangle = {\alpha' \over 2  }
\varphi(z)~~~~~~~~~~~~
\langle \bp
 x(z) x(z) \rangle = {\alpha' \over  4  }\bp
\varphi(z)
}
We  analyze the $\varphi$ dependence
 of vertices including different
combinations of background fields.
First we will consider the term $\bp V F^{-1} \p U
$. Tadpoles will vanish if
\eqn\feqn{
\p_i\p_i F^{-1} =0
}
This equation ensures also that tadpole contractions in all vertices
defining $u$ also vanish since they involve always $ F^{-1} $ or
derivatives of it.  Let us now analyze the classical dilaton
contribution
\eqn\dil{\eqalign{
 {1 \over 8 \pi } \int (-4 \varphi ) \bp \p \Phi' = &
{1 \over 8 \pi } \int (-4 \varphi ) \left\{
\p_u \p_i\Phi' ( \bp x^i F^{-1} \p U + \p x^i \bp u ) + \right. \cr
& \left. \p_i\Phi' \bp \p x^i +
\p_u^2 \Phi' F^{-1} \p U \bp u + \p_u\Phi' \p_u F^{-1} \p U \bp u +
\right. \cr
& \left. \p_u \Phi' \p_i F^{-1} \bp x^i \p U
+ \p_u \Phi' F^{-1} \p \bp U + \p_i\p_j\Phi' \bp x^i\p x^j  \right\}
}}
which implies
\eqn\constdil{ \p_i \p_j \Phi' =0~~~~~~~~~~~\p_u \p_i
\Phi' =0 }
since the terms $\p x^j \bp x^i$ or $\p x^i \bp u$ cannot
be produced from quantum corrections.
This implies that $\Phi' = -z(u) + b_i x^i$. For simplicity
we will set to zero the linear terms. Terms proportional to
$\p\bp U $ vanish on shell.

Tadpole contractions of the vertex $
\p U 2 V_i \bp x^i $, and the use of \contr, give
\eqn\aver{\eqalign{ &
  {1\over \pi \alpha'} \int d^2 z \p U 2 V_i(x,u) \p x^i
|_{tadpoles} =\cr~~~~~&
= { {1 \over 2 \pi } \int d^2 z \varphi \left\{ (\p_j^2 V_i -
\p_i \p_j V_j ) \p U \bp x^i - \p_i V_i \p \bp U -
\p_u \p_i V_i \p U \bp u \right\} }
}}
There are also terms coming from the vertex in \detother\
\eqn\dettad{\eqalign{&
 -{ 1\over \pi \alpha'} \int d^2  \bp y^a \p_u F^{-1}(x,u^{cl})
\p U y^a |_{tadpoles} = \cr~~~&
 {1 \over 2 \pi } \int d^2 z\varphi \left \{
\p_u^2 F^{-1} \p U \bp u + \p_i \p_u F^{-1} \p U \bp x^i +
\p_u F^{-1} \p \bp U \right\}
}}
Collecting all these terms together and demanding that the
coefficients of $ \p U \bp u$ and $\p U \bp x^i $ vanish we
obtain the equations
\eqn\eqnsconf{\eqalign{&
\p_u\p_i V_i - \p^2_u F^{-1} +  \p_u F^{-1} \p_u \Phi' +
F^{-1} \p_u^2 \Phi' = 0
\cr
&
\p_j\p_j V_i - \p_i \p_j V_j + \p_i  \p_u F^{-1} -
 \p_i F^{-1} \p_u \Phi'  = 0
} }
These are indeed the equations that we obtained in \equations.  One
can also have a solution with a linear dilaton $\Phi' = -z(u) + b_i
x^i $, in which case the equations \eqnsconf\ and \feqn\ acquire some
extra terms, which come from the fact that with a linear dilaton, the
one point function $\langle x \rangle$ is non zero.

This completes the argument showing conformal invariance of this
generalized sigma model.  Summarizing, the background has to satisfy
\feqn\ \constdil\ \eqnsconf.

%--------+---------+---------+---------+---------+---------+---------+
\listrefs
\bye